\documentclass[11pt, a4paper]{article}
\usepackage{geometry}
\usepackage[utf8]{inputenc}
\usepackage[T1]{fontenc}
\usepackage{palatino} 
\usepackage[english]{babel}
\usepackage{authblk} 
\usepackage{amsmath}
\usepackage{amssymb}
\usepackage{graphicx}
\usepackage{hyperref}
\RequirePackage{jabbrv}

\title{\textbf{A gridded establishment dataset as a proxy for economic activity in China}}

\author[1,2,3,*]{\small{Lei Dong}}
\author[3]{Xiaohui Yuan}
\author[3]{Meng Li}
\author[2]{Carlo Ratti}
\author[1,*]{Yu Liu}
\affil[1]{Institute of Remote Sensing and Geographical Information Systems, School of Earth and Space Sciences, Peking University, Beijing 100871, China}
\affil[2]{Senseable City Laboratory, Massachusetts Institute of Technology, Cambridge, MA 02139, USA}
\affil[3]{QuantUrban Inc., Beijing 100176, China}
\affil[*]{Corresponding authors: L.D. (arch.dongl@gmail.com) and Y.L. (liuyu@urban.pku.edu.cn)}

\date{}

\begin{document}

\maketitle

\begin{abstract}
Measuring the geographical distribution of economic activity plays a key role in scientific research and policymaking. However, previous studies and data on economic activity either have a coarse spatial resolution or cover a limited time span, and the high-resolution characteristics of socioeconomic dynamics are largely unknown. Here, we construct a dataset on the economic activity of mainland China, the gridded establishment dataset (GED), which measures the volume of establishments at a 0.01$^{\circ}$  latitude by 0.01$^{\circ}$  longitude scale. Specifically, our dataset captures the geographically based opening and closing of approximately 25.5 million firms that registered in mainland China over the period 2005-2015. The characteristics of fine granularity and long-term observability give the GED a high application value. The dataset not only allows us to quantify the spatiotemporal patterns of the establishments, urban vibrancy and socioeconomic activity, but also helps us uncover the fundamental principles underlying the dynamics of industrial and economic development.
\end{abstract}
\newpage

\section*{Background \& Summary}
An uneven geographical distribution of economic activity has a crucial impact on society \cite{storper2011regions}. It is a driving force behind a variety of social phenomena, such as population migration, industrial upgrading, and segregation. Detailed geographic data on economic activity, despite their importance, are not readily available \cite{nordhaus2006geography}. Statistical departments in a few developed countries provide aggregated establishment data to measure economic activity geographically. The Census Bureau of the United States, for example, publishes annual statistics for businesses with paid employees. The data are available by ZIP code or at coarser administrative levels \cite{cbp}. Similarly, Statistics Sweden complies a gridded Swedish socioeconomic dataset that covers all individuals and all establishments in Sweden \cite{mellander2015night-time}. However, such data are not available in China or most other countries in the world. In China, the main sources of socioeconomic data are statistical yearbooks, which only provide aggregated figures at the city level and cannot be used to analyze socioeconomic activity inside cities. Meanwhile, the comparability of the statistics from yearbooks across regions has been questioned by researchers due to inconsistent statistical standards in different regions \cite{rawski2001happening}. 

Because of the limitations of official sources, researchers often seek to use alternative data as proxies for existing measures of economic output \cite{einav2014economics}. Some `big data' sources -- e.g., nighttime lights \cite{chen2011using,henderson2012measuring}, streetview imagery \cite{naik2017computer,glaeser2018big,ma2019typeface}, mobile phone data \cite{blumenstock2015predicting,dong2017measuring}, social media data \cite{llorente2015social}, and restaurant data \cite{glaeser2017nowcasting,dong2019predicting} -- have been examined by investigating their relationships with traditional indicators of socioeconomic activity (e.g., population, GDP, and income). However, the time span covered by most big data sources is very limited; therefore, the relevant research mainly deploys cross-sectional analysis, and it is difficult to track temporal changes of a small area. Importantly, due to potential biases in the spatiotemporal coverage of big data, even with long-term observations, constructing a consistent and reliable measure is still an enormous challenge for scientists \cite{lazer2014parable,panczak2020estimating}. For instance, the number of users of different web platforms -- the main sources for big data -- are changing dynamically. It is difficult to distinguish whether the temporal changes reflected in the data are caused by changes in the number of users of the platform itself or changes in the real world. Additionally, many of these big data sources may not be publicly available or may face privacy issues. 

Under the above background, the nighttime light (NTL) dataset, a publicly available source of data covering most of the world land areas since 1992, becomes the most widely used dataset for socioeconomic activity measurement \cite{noaa,gaughan2016spatiotemporal,donaldson2016view}. However, there are still some limitations in the NTL data. First, it mainly reflects electricity-powered illumination at night \cite{mellander2015night-time}, which does not necessarily match real socioeconomic activity. Second, because of the saturation effect, NTL data can easily reach the upper limit of luminance value within cities, making it difficult to capture the variation of economic activity inside them. Third, for each grid cell, there is only the brightness value provided by NTL; however, development modes of two regions with the same luminance value (e.g., a manufacturing park and a high-tech park) may be very different. NTL cannot capture this potentially heterogeneous development. A comprehensive comparison between NTL and micro-level socioeconomic variables can be found in ref. \cite{mellander2015night-time}. In addition to NTL data, mobile phone data and social media data have also been used in recent studies to investigate economic activity \cite{llorente2015social,toole2015tracking,almaatouq2016mobile}. Although generally regarded as effective data sources, three main reasons limit their wide application in the estimation of economic activity. First, access to such data is limited. Due to privacy issues, operators or web platforms are reluctant to release their data. Second, mobile phone data or social media data mainly measure users' online social activity, which is not directly related to economic activity. Third, there are often multiple platforms in a region, so that the information reflected in mobile phone data or social media data often only represent the population covered by a particular platform, i.e., the data may not be representative \cite{e2020uncovering,grantz2020use}. 

\begin{figure}[]
    \centering
    \includegraphics[width=.9\linewidth]{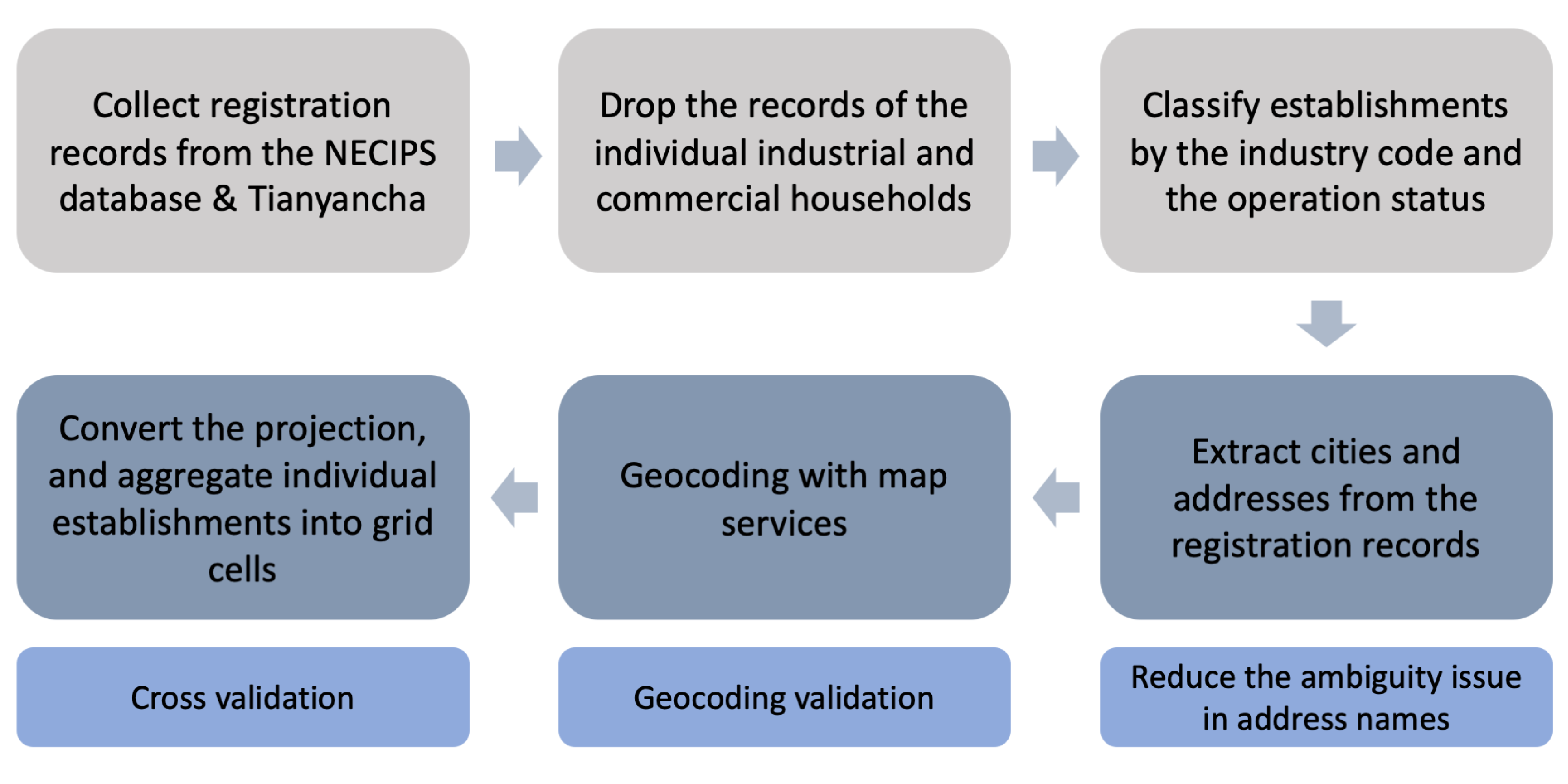}
    \caption{Schematic overview of the method used to generate the gridded establishment dataset (GED).}
    \label{fig:schematic}
\end{figure}

In this paper, using approximately 25.5 million firm registration records, we construct a geographically based dataset -- the gridded establishment dataset (GED) -- on the economic activity of mainland China. To the best of our knowledge, this is the first geocoded establishment dataset covering mainland China. An important advantage of this dataset is that it simultaneously provides high spatial resolution (0.01$^{\circ}$  latitude by 0.01$^{\circ}$  longitude, approximately 1.1 km $\times$ 1.1 km at the equator) and long-duration observations (2005-2015, 11 years). For each grid cell, we also provide the number of establishments by industry, which allows us to analyze the heterogeneous development trajectories of a region. The workflow to construct the dataset is shown in Fig.~\ref{fig:schematic} and detailed in the Methods section.

To validate this dataset, we analyze the accuracy of the geocoding process and compare the GED with several socioeconomic indicators and NTL data at the city level. The results show that the GED can effectively reflect socioeconomic activity and achieves a better performance in terms of fit than that of the NTL. The characteristics of fine granularity and long-term observability give this dataset high application values. It not only can allow researchers and policymakers to quantify the spatiotemporal patterns of economic activity but also can help us uncover the fundamental principles underlying the dynamics of industrial development. Additionally, having a greater availability of these granular spatiotemporal data can advance the potential of machine learning models for urban studies. For example, the GED can serve as tags for satellite/streetview imagery to build machine learning models to predict socioeconomic dynamics \cite{jean2016combining}.

Although the GED is a major step towards measuring the geographical distribution of economic activity in China, we should note that it also has several limitations. First, we only have the registered addresses of firms, which may not be the same as the operational addresses. Especially for some industrial parks, due to some location-based policies, e.g. tax subsidies, some firms may register within these industrial parks but operate elsewhere. Second, because of the change in place names, there may be errors in the geocoding process for historical addresses. Therefore, we limit the starting date of the dataset to 2005, when Amap and Baidu, the location-based service providers we use to geocode the address, began to provide online map services in China. Third, business size (e.g., number of employees or revenue-related information) is undocumented in the original data; thus, we can only use establishment counts to represent economic activity. 

\section*{Methods}
\subsection*{Data source and preprocessing}

Each newly established firm in mainland China needs to register at the local Administration for Industry and Commerce (AIC) by providing an array of detailed information, such as firm name, address, industry classification, and stockholders. After approval, the relevant information is publicized on the Internet, and everyone can retrieve these data from an online system named the National Enterprise Credit Information Publicity System (NECIPS, \url{http://www.gsxt.gov.cn/index.html}). Our original data are collected from this system via web scraping. Specifically, we collect the names of firm that were established between 2005 and 2015 from \textit{Tianyancha} \cite{tianyancha}, a website that provides comprehensive firm information in China. By querying firm names, we then obtain the detail page of each firm. Finally, we collect 83.7 million registration records.

To reduce potential noise in the original data, we perform preprocessing as follows: 1) We drop the record of ``individual industrial and commercial household'' (57.8 million records), a business type whose scale is generally small -- fewer than seven employees by law. Another reason for this filter is that the registered addresses of ``individual industrial and commercial households'' are often at village or township level, and there is no clear street address to perform geocoding. 2) According to the operational status in the year (2018) of data collection, we classify firms into three categories: \textit{existing}, \textit{cancelled}, and \textit{other}, see the preprocessing code for details. Here, \textit{existing} means that the firm has been operating from the time of registration until (at least) 2018; \textit{cancelled} means that the firm was closed at a certain time between the registration year and 2018; \textit{other} represents the remaining status, such as unknown status and moving out. Note that this classification does not affect the number of establishments in the statistics every year, but only adds a dimension to evaluate the operation status of establishments. 3) According to the industry classification, we further label establishments with tags of primary, secondary, and tertiary sectors (Table~\ref{tab:1}). After data preprocessing, there are a total of 25,545,850 establishments left.

Figure~\ref{fig:basic}a show the number of newly established firms by industry each year. To demonstrate the representativeness of this dataset, we further collect the official statistics based on a report published by the State Administration for Industry and Commerce \cite{AIC}. Specifically, according to the report, the number of newly established firms in 2015 was about 12,200 per day, or 4.45 million per year (the dashed line in Fig.~\ref{fig:basic}a). In our data, there were 4.85 million establishments in 2015, accounting for 109.2\% of the government released figures (Fig.~\ref{fig:basic}b). The coverage ratio greater than 100\% is probably because the government has subtracted the firms that were cancelled during the registration year when counting the yearly establishments. Overall, our results are identical with those of official statistics (Fig.~\ref{fig:basic}b).

\begin{table}[]
    \centering
    \begin{tabular}{*2l}
     \hline
     \hline
     Industry & Subgroup \\
     \hline
     Primary & Agriculture; forestry; husbandry; fishery \\
     \hline
     Secondary &  Mining; manufacturing; construction\\
     & Production and supply of electricity, water and gas \\
     \hline
     Tertiary & Transport, storage and post services \\
      & Accommodation and catering services \\
      & Leasing and business services; education; finance \\
      & Scientific research and technical services \\ 
      & Wholesale and retail; real estate\\
      & Culture, sports and entertainment\\ 
      & Health and social work \\
      & Resident services, maintenance and other service \\
      & Public management, social security, and social organization \\
      & Information transmission, software, and IT services\\
      & Water conservancy, environment, and public facilities management\\
     \hline
     \hline
    \end{tabular}
    \caption{Industry classification}
    \label{tab:1}
\end{table}

\subsection*{Geocoding}

Geocoding is the key step for building a geolocated dataset \cite{de2019geocoding}. To obtain high-quality geocoding results, it is important to know the city to which the firm address belongs. The city and address can then be passed into the online map application program interface (API) to ensure that the search results belong to the input city, which can significantly reduce the problem of ambiguity in place names. In other words, cities may have streets or buildings with the same name, and only querying the street name may return multiple coordinates, for example, one in city $A$ and one in city $B$, making it difficult to match the address with the right coordinate. We notice that a considerable proportion of the city data is missing or inaccurate in the NECIPS database. To solve the missing value issue and obtain the coordinates of each firm, we have taken the following steps:

\begin{itemize}
    \item We build a city list containing the full name and short name of all prefecture-level cities (including \textit{dijishi}, \textit{diqu}, \textit{meng}, and \textit{autonomous prefecture}) and counties (including \textit{shixiaqu}, \textit{xian}, \textit{qi}, and \textit{autonomous county}) in mainland China, see ref. \cite{citydefination} for details on the administrative divisions of China. 
    \item We inspect the city list; if a city name (full name) appears in the address of a registration record, the firm is then classified into this city. For example, the address `Beijing City, Haidian District, Zhongguancun Street' belongs to Beijing City.
    \item For cities where the full name cannot be found, we check whether the short name of the city appears in the address. For example, `Shanghai, Pudong District, and Haiyang Building’ belong to Shanghai City because `Shanghai' is a short name of `Shanghai City.'
    \item For cities that can not be matched to any cities, we then inspect the county/district names. For example, the address `Haidian District, Zhongguancun Street' belongs to Beijing City, as `Haidian' is a district of Beijing.
    \item If the address field does not contain any description of the city/county/district name, we try to extract the city information from other fields such as the registration authority. In the end, the corresponding cities or countries could not be found for approximately 3\% of the total records. The step of extracting city names can be processed automatically by some Python packages, e.g., the `cpca' package \cite{cpca}. 
    \item After having the city and address information, we geocode them with AMap API \cite{gaode}, and obtain the coordinate information (longitude, latitude). Because the registered addresses will be checked by AIC, it is unlikely that there would be typo issues. AMap also automatically handles typo issues according to the document of the API \cite{gaode}. 
    \item Finally, we convert all geocoded coordinates from the GCJ02 projection, a special projection system in China \cite{coord}, into the world geodetic system 1984 (WGS84) projection, and then aggregate the coordinates according to a 0.01$^{\circ}$ latitude by 0.01$^{\circ}$ longitude grid.
\end{itemize}

\begin{figure}[]
    \centering
    \includegraphics[width=1.\linewidth]{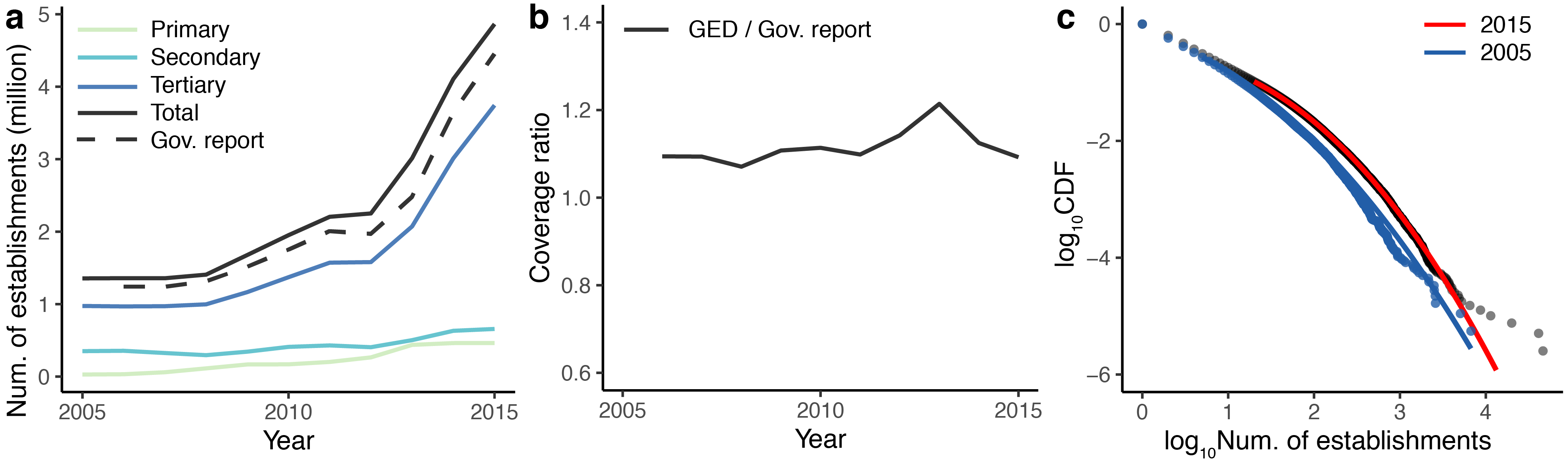}
    \caption{Basic statistics of the GED. (\textbf{a}) Number of establishments in different sectors each year. (\textbf{b}) The coverage ratio of the GED. (\textbf{c}) The cumulative distribution function (CDF) of establishments at the grid cell level. The log-normal fitting lines are estimated by the poweRlaw package in R \cite{gillespie2015fitting}.}
    \label{fig:basic}
\end{figure}

\begin{figure}[]
    \centering
    \includegraphics[width=1.\linewidth]{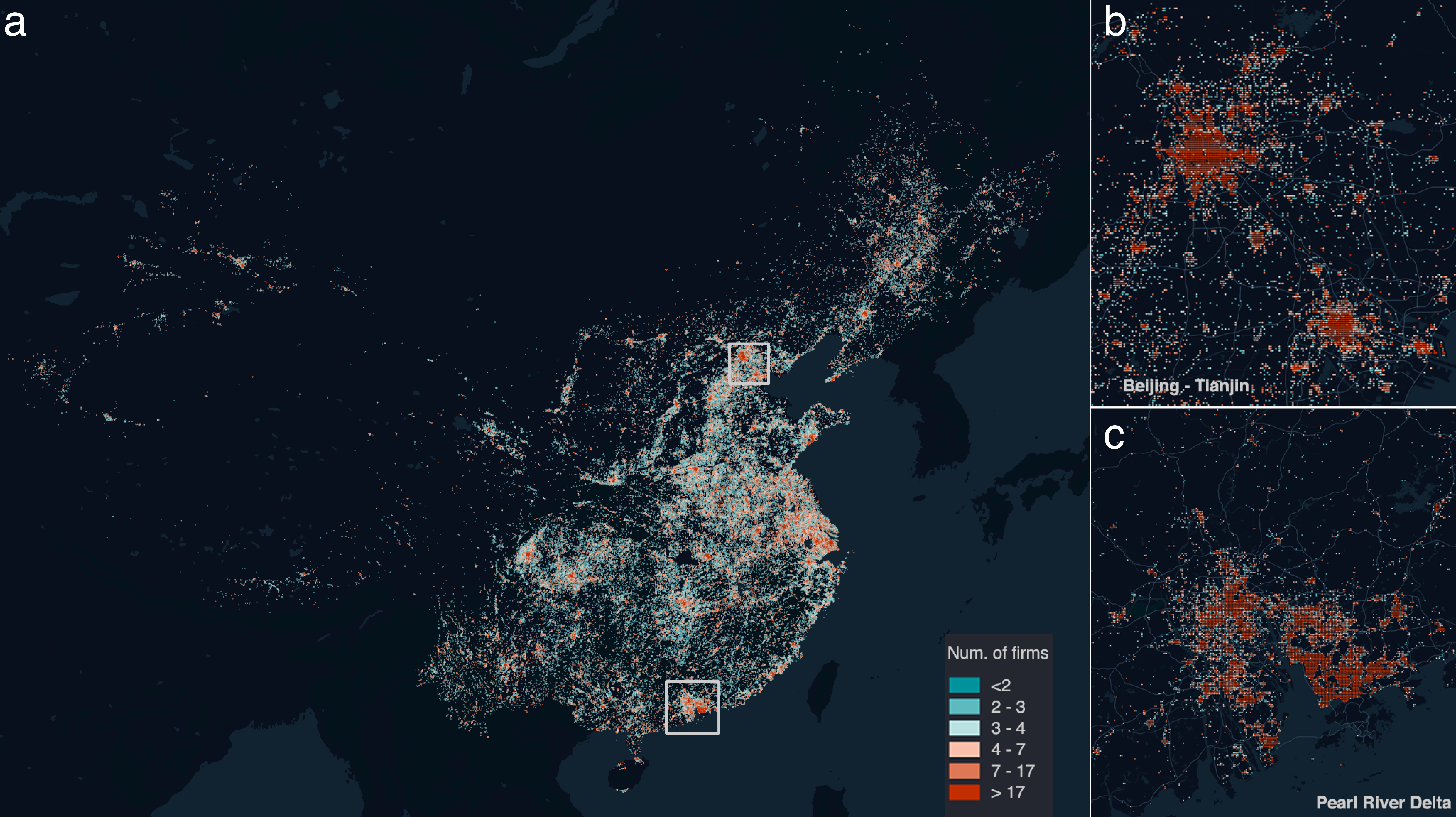}
    \caption{Geographical distribution of establishments (2015). (\textbf{a}) All of mainland China. (\textbf{b}) Beijing-Tianjian-Hebei areas. (\textbf{c}) Pearl River Delta. The South China Sea Islands are not fully shown on the map. (Base map: Mapbox. Visualization tool: Kepter.gl.)}
    \label{fig:map}
\end{figure}

As shown in Fig.~\ref{fig:basic}b, we find that the distributions of the grid-cell values are well fitted by a log-normal function, which is in line with most socioeconomic characteristics \cite{limpert2001log}. Meanwhile, we plot the map of newly established firms in 2015 (Fig.~\ref{fig:map}). As seen from the map, our dataset clearly shows the geographic pattern of economic activity in China -- the vast majority of economic activity is concentrated in the eastern coastal areas, especially the three megacity clusters: Beijing-Tianjin-Hebei, the Yangtze River Delta, and the Pearl River Delta. Unlike the NTL data, the establishment dataset does not have the saturation effect, so it can easily capture variations within cities (Figs.~\ref{fig:map}b,c).

\section*{Data Records}
Data are available at the figshare repository \cite{figshare}. The repository contains four data files, and the variable descriptions of the main file (`GED\_2005-2015\_v2.csv') are reported in Table 2. Each row of the file represents the record of one grid cell, including the centroid longitude and latitude, year of the record, total number of establishments, number of establishments in different operational statuses, and number of establishments in different industries (Table~\ref{tab:2}). It should be noted that for a cancelled firm, the date variable here represents the year when the firm was registered, not the year when it was cancelled.

To reduce the file size, for a given year if a grid cell does not have any registration record, this grid cell is not included in the dataset. For ease of use, we also link the coordinates of cells with the county, city and province information. The basic statistical summary at the province level is shown in Table~\ref{tab:3}. Guangdong, Jiangsu, Shangdong, Zhejiang and Shanghai rank among the top five in terms of number of establishments. The reason why Shandong ranks high is because of its large population, and the other four provinces are recognized as the regions with the strongest entrepreneurial spirit in China. 

We also provide three-year -- 2005, 2010, and 2015 -- unaggregated data to facilitate researchers to conduct research on different scales. For the data description of the unaggregated data, see the `read-me' file under the figshare repository path. 

\begin{table}[]
    \centering
    \begin{tabular}{*2l}
     \hline
     \hline
     Variable name & Description \\
     \hline
     idx & Index of the grid cell\\
     lon & Longitude of the center of the grid cell (WGS84 projection)\\
     lat & Latitude of the the center of the grid cell (WGS84 projection)\\
     date & Year of the establish record \\
     establish\_total & Total number of establishments within the cell\\
     - establish\_prim & - Number of newly established firms in primary industry\\
     - establish\_seco & - Number of newly established firms in secondary industry\\
     - establish\_tert & - Number of newly established firms in tertiary industry\\
     exist\_total & Number of existing firms\\
     - exist\_prim & - Number of existing firms in primary industry\\
     - exist\_seco & - Number of existing firms in secondary industry\\
     - exist\_tert & - Number of existing firms in tertiary industry\\
     cancel\_total & Number of cancelled firms\\
     - cancel\_prim & - Number of cancelled firms in primary industry\\
     - cancel\_seco & - Number of cancelled firms in secondary industry\\
     - cancel\_tert & - Number of cancelled firms in tertiary industry\\
     other\_total & Number of establishments with unknown operation status\\
     county & Name of the county-level division where the grid cell is located\\
     city & Name of the prefecture city where the grid cell is located\\
     province & Name of the province where the grid cell is located\\
     \hline
     \hline
    \end{tabular}
    \caption{List of variables}
    \label{tab:2}
\end{table}

\begin{table}[]
    \centering
    \begin{tabular}{*7l}
     \hline
     \hline
     Province & $N_{total}$ & $Min$ & $Max$ & $Mean$ & $SD$ \\
     \hline
Anhui&812,773&1&1,275&6.73&20.23\\
Beijing&1,356,979&1&4,758&30.48&104.11\\
Chongqing&651,720&1&1,409&9.30&34.55\\
Fujian&809,078&1&2,526&8.97&34.67\\
Gansu&304,597&1&1,718&5.63&19.94\\
Guangdong&3,004,904&1&41,232&18.45&134.45\\
Guangxi&629,717&1&1,642&8.39&30.40\\
Guizhou&551,186&1&2,000&7.28&31.97\\
Hainan&184,424&1&1,714&11.90&49.48\\
Hebei&1,017,968&1&938&5.27&16.36\\
Heilongjiang&460,459&1&782&6.28&17.07\\
Henan&1,101,598&1&1,172&5.72&20.22\\
Hubei&952,163&1&1,261&7.65&27.60\\
Hunan&635,266&1&860&5.54&19.51\\
Innter Mongolia&410,483&1&474&6.31&15.15\\
Jiangsu&2,635,874&1&46,388&10.49&104.50\\
Jiangxi&547,476&1&950&6.25&17.78\\
Jilin&422,966&1&11,412&7.11&62.65\\
Liaoning&726,469&1&685&7.15&20.37\\
Ningxia&127,697&1&643&7.58&21.55\\
Qinghai&83,245&1&450&5.28&15.18\\
Shaanxi&79,774&1&992&2.46&11.70\\
Shandong&1,840,092&1&1,238&6.21&19.90\\
Shanghai&1,607,515&1&6,724&40.72&142.14\\
Shanxi&515,799&1&819&4.29&13.51\\
Sichuan&1,044,063&1&3,383&7.06&36.84\\
Tianjin&392,245&1&3,500&12.97&45.23\\
Tibet&37,346&1&489&5.29&16.63\\
Xinjiang&306,840&1&650&7.01&20.04\\
Yunnan&519,001&1&1,394&7.43&26.84\\
Zhejiang&1,734,792&1&2,109&8.46&25.01\\
     \hline
     \hline
    \end{tabular}
    \caption{Statistical summary of the establishments (2005-2015). $N_{total}$ is the total number of establishments within the province. $Min$, $Max$, $Mean$, and $SD$ (the standard deviation) are the statistical indicators at the grid cell level.}
    \label{tab:3}
\end{table}

\section*{Technical Validation}

To validate the dataset, we analyze the accuracy of geocoding and the correlation between our dataset and other socioeconomic indicators.

\subsection*{Validation of geocoding}
According to city size and geographical location, we select three representative cities and randomly sample 10,000 records for analysis and comparison. The three cities are Beijing (large city), Changsha (medium-sized city), and Anshan (small city). For geocoding accuracy, we use two methods to evaluate. The first is the spatial resolution returned by the Amap API. This is the geographical level that the map can match according to the input address. The finer the geographical resolution, the higher the accuracy may be. As shown in Table~\ref{tab:4}, the finest match is to the street number, followed by the point of interest (POI). Any address that can match these two levels is generally of high accuracy, and we find that more addresses are matched at these two levels in large cities than in medium and small cities: 88\% of Beijing compared with 77\% of Changsha and 48\% of Anshan. Moreover, if we consider two more levels, to the level of villages and towns, then the geocoding accuracy of these three cities is 96\%, 95\% and 84\%, respectively.

Another verification method is to use the `confidence indicator' returned by Baidu Maps. (Amap and Google Maps do not provide this measure, but considering Amap and Baidu Maps provide similar services in China, therefore we use the `confidence indicator' obtained from Baidu Maps as a benchmark to evaluate the quality of the address attributes of firms.) According to the document of Baidu Map \cite{baidu}, geocoding confidence $\geq$ 50 means the absolute geocoding error is less than 1000m. We use Baidu Maps to re-geocode these sampled addresses, the results show that the proportions of coordinate errors less than 1000m, i.e. confidence $\geq$ 50, in the three cities are 93.1\%, 92.2\% and 74.1\%, respectively, implying that the accuracy of the geocoding services is significantly higher in large cities than in small cities.

\begin{table}[]
    \centering
    \begin{tabular}{*4c}
     \hline
     \hline
     Spatial resolution & Beijing & Changsha & Anshan \\
     \hline
     Street number & 7,108 & 4,673 & 3,475 \\
     POI & 1,665 & 3,005 & 1,276  \\
     Village & 537 & 1,387 & 2,500  \\
     Town & 313 & 431 & 1,128  \\
     Road & 249 & 347 & 962  \\
     Other & 128 & 157 & 659  \\
     \hline
     \hline
    \end{tabular}
    \caption{Geocoding resolution}
    \label{tab:4}
\end{table}

\subsection*{Validation with city socioeconomic indicators}

\begin{figure}[]
    \centering
    \includegraphics[width=1.\linewidth]{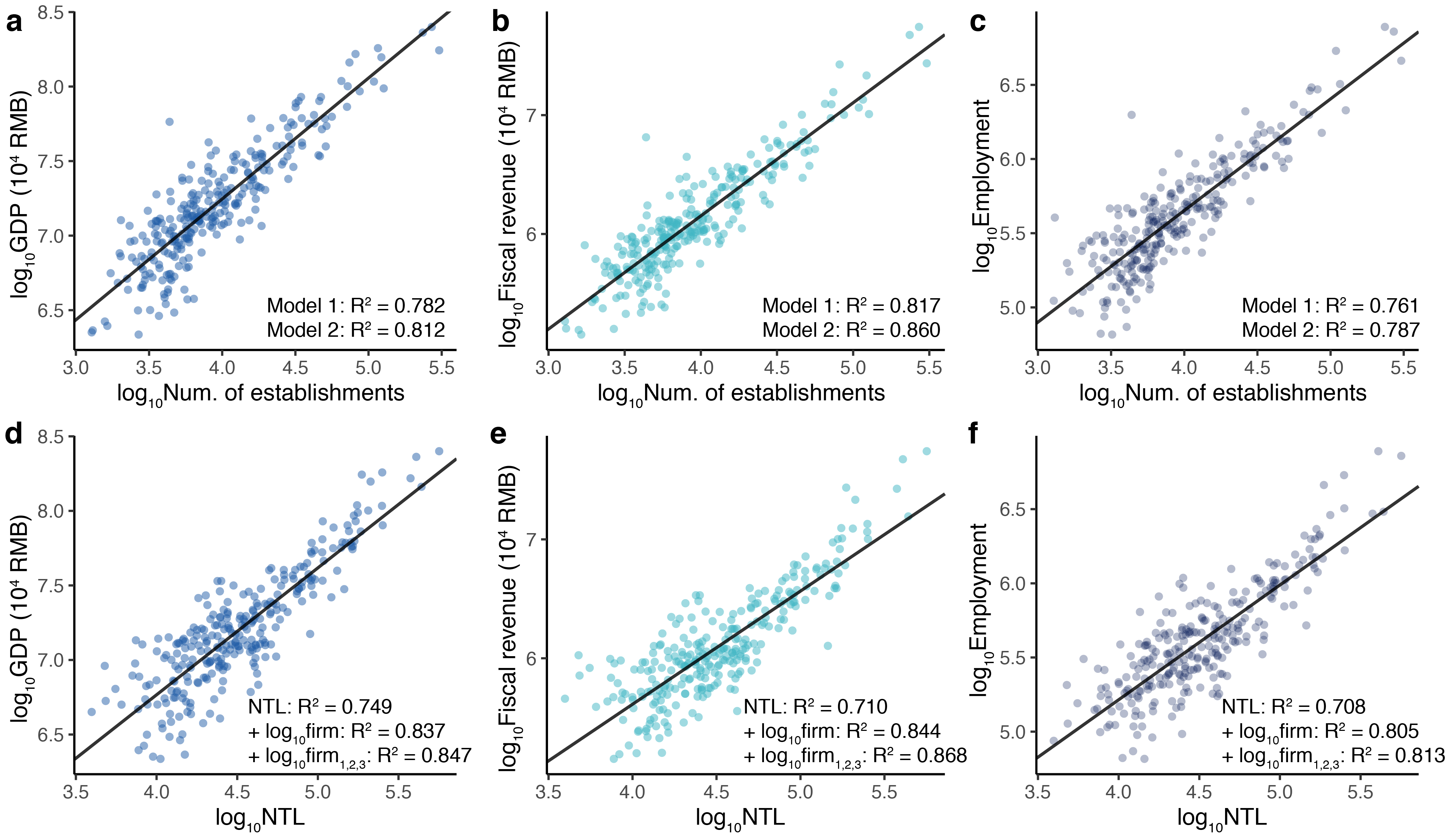}
    \caption{Validation with socioeconomic indicators at city level. (\textbf{a, b, c}) Correlation between number of establishments and (\textbf{a}) GDP, (\textbf{b}) fiscal revenue, and (\textbf{c}) employment. (\textbf{d, e, f}) Similar to (\textbf{a-c}) but replacing establishments with the nighttime light data. We also add $\log_{10}$NTL in Eq.(1) and (2), and present the regression results in (\textbf{d-f}).}
    \label{fig:validation}
\end{figure}

To further justify the validity of the data, we aggregate the number of newly established firms at the city level, and compare it with the city's socioeconomic indicators: GDP, fiscal revenue, and employment. These indicators are collected from the city statistical yearbook, and some cities are excluded due to a lack of statistical data from the yearbook. We present the results of two fitting models in Fig.~\ref{fig:validation}. One is a linear regression between the total number of establishments and the variables of interest $y$, denoted as Model 1:

\begin{equation}
    \log_{10}(y) = \beta_0 + \beta_1\log_{10}(\text{Est}) + \epsilon
\end{equation}

The second model decomposes the number of establishments into three sectors, i.e., the primary, secondary, and tertiary sectors. Model 2 is as follows:

\begin{equation}
    \log_{10}(y) = \beta_0 + \beta_1\log_{10}(\text{Est}_1) + \beta_2\log_{10}(\text{Est}_2) + \beta_3\log_{10}(\text{Est}_3) +\epsilon
\end{equation}

To make a comparison with the nighttime light data, we also collect the Visible Infrared Imaging Radiometer Suite (VIIRS) NTL data for 2015 (available through \url{www.ngdc.noaa.gov/eog/viirs}). We use the annual `vcm-orm-ntl' average radiance layer, which has undergone an outlier removal process (exclusion of data impacted by stray light, lightning, cloud cover, etc.). Similar to the establishment data, we aggregate the values of nighttime light data at the city level. To make the results comparable, we used the 2015 data of each dataset for comparison.

Figure~\ref{fig:validation} shows the fitting results (adjusted $R^2$) of GDP, fiscal revenue, and employment at the city level. In general, both our data and the nighttime light data perform well in estimating these socioeconomic variables, with adjusted $R^2$ values over 0.70 in all cases. However, the accuracy of the establishment data is better than that of the the nighttime light data in predicting all three variables (Model 1), especially for fiscal revenue (0.817 vs. 0.710) and employment (0.761 vs. 0.708), and decomposing establishments into three sectors, i.e., Model 2, further improves the adjusted $R^2$ (Fig.~\ref{fig:validation}a-c). Additionally, we add $\log_{10}$NTL in Eq.(1) and (2) to test how much additional variance can be explained by the combined model (Fig.~\ref{fig:validation}d-f). Results demonstrate that the combined version of Eq.(2) achieves the highest predicting accuracy, implying that the establishment data and NTL data respectively capture some elements of the socioeconomic indicators, thus they can be jointly used in practice.

\begin{figure}[]
    \centering
    \includegraphics[width=1.\linewidth]{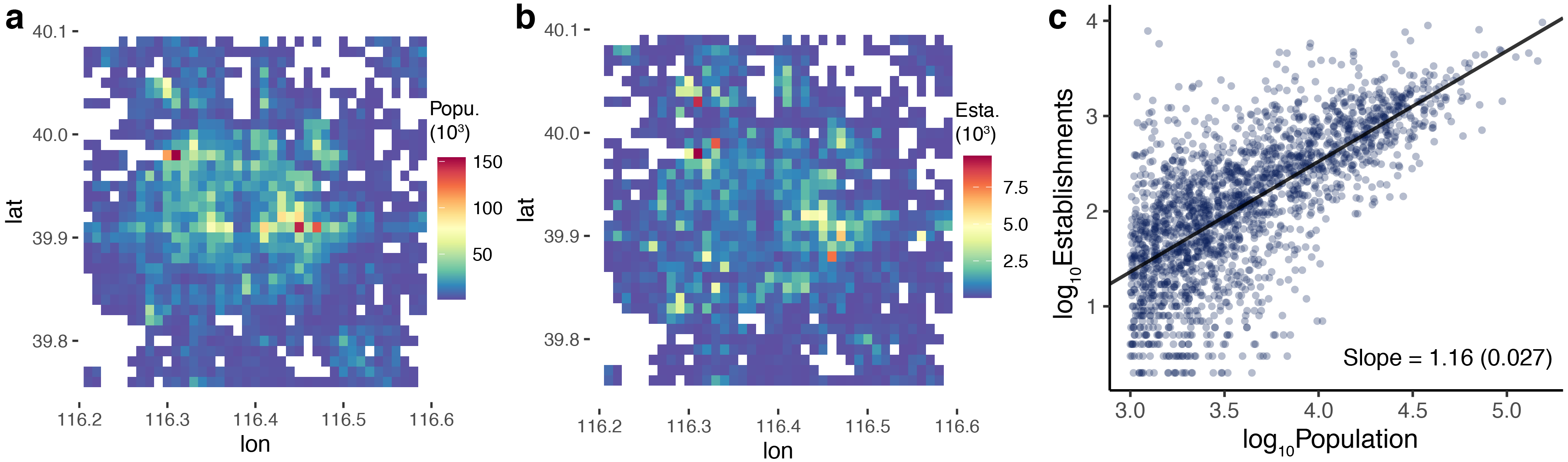}
    \caption{Validation with urban scaling theory. (\textbf{a, b}) Geographical distributions of (\textbf{a}) population and (\textbf{b}) establishments. (\textbf{c}) Power-law relationship between population and establishments. The exponent $\alpha = 1.16$ is greater than 1.}
    \label{fig:scaling}
\end{figure}

Finally, we validate our dataset by linking it with a granular population dataset to verify the scaling law of cities. As demonstrated in numerous studies \cite{bettencourt2007growth,li2017simple}, urban socioeconomic activity $Y$ scales with population $P$ in a simple power-law manner: $Y \sim P^{\alpha}$, where $\alpha$ is called a scaling exponent (or elasticity in the term of economics). To test the power-law relationship within cities, we collect a mobile phone estimated population dataset from ref. \cite{dong2020understanding}. We match the population with the establishment data under the same spatial resolution (0.01$^{\circ}$  latitude by 0.01$^{\circ}$ longitude, see Fig.~\ref{fig:scaling}), and then run a simple regression to estimate the scaling coefficient $\alpha$ of Beijing. According to the assumption of the scaling theory, we constrain the analysis within the urbanized area by setting a population density threshold of 1,000 people/$\text{km}^2$, and drop cells with population density lower than this threshold. Result shows that the power-law relationship between population and establishments holds well within cities, and the scaling exponent $\alpha = 1.16$ is greater than 1, which is very close to the theoretical numbers derived from different models \cite{bettencourt2013origins}.

\section*{Usage Notes}
The GED can be used in geographic information systems (GIS), such as ArcGIS and QGIS, as well as statistical software, such as Python, R, MATLAB, and Stata. In GIS software, the dataset can be imported as a vector layer or a raster layer. To match with other geographical datasets, such as industrial park boundaries~\cite{zheng2017birth} or water/air quality monitoring records~\cite{he2020watering}, users can apply the spatial join function in GIS software by joining attributes from GED to another data based on the spatial relationship \cite{spatialjoin}. For example, we can draw a buffer for each industrial park, and then extract the grid cells and the corresponding establishments that fall within this buffer. If the resolution of the GED is inconsistent with other data sources in use, some resampling methods used in remote sensing areas could also be applied to this dataset \cite{lyons2018comparison}. 

Additionally, the county/city/province name provided in this dataset is based on the administrative boundary of 2019. Therefore, if the GED is matched with other statistical data by couty/city/province name, the impact of changes in the name (and the corresponding spatial range) should be considered. It's also notable that the grid cell index of GED is under WGS84 coordinates. Thus, the spherical area of grid cells in different regions is different (approximately 1.1 km  $\times$ 1.1 km at the equator and 0.85 km $\times$ 1.1 km at Beijing). 

Finally, as mentioned in the introduction section, due to some location-based policies, there will be a large number of registered enterprises in some specific areas, such as the Shanghai Free Trade Zone. The data points on the far right of Fig.~\ref{fig:basic}c indicate that there are more than 10,000 establishments in each of these grid cells. Such spatial `outliers' would have a certain impact on some micro-scale analyses. To address this issue, we suggest taking an upper limit on the value in each cell or useing the logarithmic transformation of the data when performing micro-scale analyses.

\subsection*{Code availability}
The preprocess script, validation dataset and the R code that performs the statistical analysis are available through \url{https://github.com/quanturban/firm}.

\section*{Acknowledgements}
We thank Rui Du and Siqi Zheng for helpful comments. L.D. was supported by the National Natural Science Foundation of China (no. 41801299) and the China Postdoctoral Science Foundation (2018M630026).

\section*{Author contributions}
L.D., X.H., C.R. and Y.L. designed research; L.D. performed research; L.D. and M.L. analyzed data; L.D. wrote the paper. All authors reviewed the paper.

\section*{Competing interests}
L.D. and X.Y. are co-founders of QuantUrban, a company that uses big data and machine learning to measure socioeconomic dynamics of cities.


\begin{thebibliography}{10}
\urlstyle{rm}
\expandafter\ifx\csname url\endcsname\relax
  \def\url#1{\texttt{#1}}\fi
\expandafter\ifx\csname urlprefix\endcsname\relax\def\urlprefix{URL }\fi
\expandafter\ifx\csname doiprefix\endcsname\relax\def\doiprefix{DOI: }\fi
\providecommand{\bibinfo}[2]{#2}
\providecommand{\eprint}[2][]{\url{#2}}

\bibitem{storper2011regions}
\bibinfo{author}{Storper, M.}
\newblock \bibinfo{journal}{\bibinfo{title}{Why do regions develop and change?
  The challenge for geography and economics}}.
\newblock {\emph{\JournalTitle{Journal of Economic Geography}}}
  \textbf{\bibinfo{volume}{11}}, \bibinfo{pages}{333--346}
  (\bibinfo{year}{2011}).

\bibitem{nordhaus2006geography}
\bibinfo{author}{Nordhaus, W.~D.}
\newblock \bibinfo{journal}{\bibinfo{title}{Geography and macroeconomics: New
  data and new findings}}.
\newblock {\emph{\JournalTitle{Proceedings of the National Academy of
  Sciences}}} \textbf{\bibinfo{volume}{103}}, \bibinfo{pages}{3510--3517}
  (\bibinfo{year}{2006}).

\bibitem{cbp}
\bibinfo{author}{{Census Bureau}}.
\newblock \bibinfo{title}{County business patterns}.
\newblock
  \bibinfo{howpublished}{\url{https://www.census.gov/programs-surveys/cbp.html}}
  (\bibinfo{year}{2018}).

\bibitem{mellander2015night-time}
\bibinfo{author}{Mellander, C.}, \bibinfo{author}{Lobo, J.},
  \bibinfo{author}{Stolarick, K.} \& \bibinfo{author}{Matheson, Z.}
\newblock \bibinfo{journal}{\bibinfo{title}{Night-time light data: A good proxy
  measure for economic activity?}}
\newblock {\emph{\JournalTitle{PLoS One}}} \textbf{\bibinfo{volume}{10}},
  \bibinfo{pages}{e0139779} (\bibinfo{year}{2015}).

\bibitem{rawski2001happening}
\bibinfo{author}{Rawski, T.~G.}
\newblock \bibinfo{journal}{\bibinfo{title}{What is happening to {China's GDP}
  statistics?}}
\newblock {\emph{\JournalTitle{China Economic Review}}}
  \textbf{\bibinfo{volume}{12}}, \bibinfo{pages}{347--354}
  (\bibinfo{year}{2001}).

\bibitem{einav2014economics}
\bibinfo{author}{Einav, L.} \& \bibinfo{author}{Levin, J.}
\newblock \bibinfo{journal}{\bibinfo{title}{Economics in the age of big data}}.
\newblock {\emph{\JournalTitle{Science}}} \textbf{\bibinfo{volume}{346}},
  \bibinfo{pages}{1243089} (\bibinfo{year}{2014}).

\bibitem{chen2011using}
\bibinfo{author}{Chen, X.} \& \bibinfo{author}{Nordhaus, W.~D.}
\newblock \bibinfo{journal}{\bibinfo{title}{Using luminosity data as a proxy
  for economic statistics}}.
\newblock {\emph{\JournalTitle{Proceedings of the National Academy of
  Sciences}}} \textbf{\bibinfo{volume}{108}}, \bibinfo{pages}{8589--8594}
  (\bibinfo{year}{2011}).

\bibitem{henderson2012measuring}
\bibinfo{author}{Henderson, J.~V.}, \bibinfo{author}{Storeygard, A.} \&
  \bibinfo{author}{Weil, D.~N.}
\newblock \bibinfo{journal}{\bibinfo{title}{Measuring economic growth from
  outer space}}.
\newblock {\emph{\JournalTitle{American Economic Review}}}
  \textbf{\bibinfo{volume}{102}}, \bibinfo{pages}{994--1028}
  (\bibinfo{year}{2012}).

\bibitem{naik2017computer}
\bibinfo{author}{Naik, N.}, \bibinfo{author}{Kominers, S.~D.},
  \bibinfo{author}{Raskar, R.}, \bibinfo{author}{Glaeser, E.~L.} \&
  \bibinfo{author}{Hidalgo, C.~A.}
\newblock \bibinfo{journal}{\bibinfo{title}{Computer vision uncovers predictors
  of physical urban change}}.
\newblock {\emph{\JournalTitle{Proceedings of the National Academy of
  Sciences}}} \textbf{\bibinfo{volume}{114}}, \bibinfo{pages}{7571--7576}
  (\bibinfo{year}{2017}).

\bibitem{glaeser2018big}
\bibinfo{author}{Glaeser, E.~L.}, \bibinfo{author}{Kominers, S.~D.},
  \bibinfo{author}{Luca, M.} \& \bibinfo{author}{Naik, N.}
\newblock \bibinfo{journal}{\bibinfo{title}{Big data and big cities: The
  promises and limitations of improved measures of urban life}}.
\newblock {\emph{\JournalTitle{Economic Inquiry}}}
  \textbf{\bibinfo{volume}{56}}, \bibinfo{pages}{114--137}
  (\bibinfo{year}{2018}).

\bibitem{ma2019typeface}
\bibinfo{author}{Ma, R.}, \bibinfo{author}{Wang, W.}, \bibinfo{author}{Zhang,
  F.}, \bibinfo{author}{Shim, K.} \& \bibinfo{author}{Ratti, C.}
\newblock \bibinfo{journal}{\bibinfo{title}{Typeface reveals spatial economical
  patterns}}.
\newblock {\emph{\JournalTitle{Scientific Reports}}}
  \textbf{\bibinfo{volume}{9}}, \bibinfo{pages}{1--9} (\bibinfo{year}{2019}).

\bibitem{blumenstock2015predicting}
\bibinfo{author}{Blumenstock, J.}, \bibinfo{author}{Cadamuro, G.} \&
  \bibinfo{author}{On, R.}
\newblock \bibinfo{journal}{\bibinfo{title}{Predicting poverty and wealth from
  mobile phone metadata}}.
\newblock {\emph{\JournalTitle{Science}}} \textbf{\bibinfo{volume}{350}},
  \bibinfo{pages}{1073--1076} (\bibinfo{year}{2015}).

\bibitem{dong2017measuring}
\bibinfo{author}{Dong, L.} \emph{et~al.}
\newblock \bibinfo{journal}{\bibinfo{title}{Measuring economic activity in
  {C}hina with mobile big data}}.
\newblock {\emph{\JournalTitle{EPJ Data Science}}}
  \textbf{\bibinfo{volume}{6}}, \bibinfo{pages}{29} (\bibinfo{year}{2017}).

\bibitem{llorente2015social}
\bibinfo{author}{Llorente, A.}, \bibinfo{author}{Garcia-Herranz, M.},
  \bibinfo{author}{Cebrian, M.} \& \bibinfo{author}{Moro, E.}
\newblock \bibinfo{journal}{\bibinfo{title}{Social media fingerprints of
  unemployment}}.
\newblock {\emph{\JournalTitle{PloS One}}} \textbf{\bibinfo{volume}{10}},
  \bibinfo{pages}{e0128692} (\bibinfo{year}{2015}).

\bibitem{glaeser2017nowcasting}
\bibinfo{author}{Glaeser, E.~L.}, \bibinfo{author}{Kim, H.} \&
  \bibinfo{author}{Luca, M.}
\newblock \bibinfo{title}{Nowcasting the local economy: Using yelp data to
  measure economic activity}.
\newblock \bibinfo{type}{Working Paper} \bibinfo{number}{24010},
  \bibinfo{institution}{National Bureau of Economic Research}
  (\bibinfo{year}{2017}).
\newblock \url{10.3386/w24010}.

\bibitem{dong2019predicting}
\bibinfo{author}{Dong, L.}, \bibinfo{author}{Ratti, C.} \&
  \bibinfo{author}{Zheng, S.}
\newblock \bibinfo{journal}{\bibinfo{title}{Predicting neighborhoods’
  socioeconomic attributes using restaurant data}}.
\newblock {\emph{\JournalTitle{Proceedings of the National Academy of
  Sciences}}} \textbf{\bibinfo{volume}{116}}, \bibinfo{pages}{15447--15452}
  (\bibinfo{year}{2019}).

\bibitem{lazer2014parable}
\bibinfo{author}{Lazer, D.}, \bibinfo{author}{Kennedy, R.},
  \bibinfo{author}{King, G.} \& \bibinfo{author}{Vespignani, A.}
\newblock \bibinfo{journal}{\bibinfo{title}{The parable of {Google flu}: traps
  in big data analysis}}.
\newblock {\emph{\JournalTitle{Science}}} \textbf{\bibinfo{volume}{343}},
  \bibinfo{pages}{1203--1205} (\bibinfo{year}{2014}).

\bibitem{panczak2020estimating}
\bibinfo{author}{Panczak, R.}, \bibinfo{author}{Charles-Edwards, E.} \&
  \bibinfo{author}{Corcoran, J.}
\newblock \bibinfo{journal}{\bibinfo{title}{Estimating temporary populations: a
  systematic review of the empirical literature}}.
\newblock {\emph{\JournalTitle{Palgrave Communications}}}
  \textbf{\bibinfo{volume}{6}}, \bibinfo{pages}{1--10} (\bibinfo{year}{2020}).

\bibitem{noaa}
\bibinfo{author}{NOAA}.
\newblock \bibinfo{title}{Version 4 dmsp-ols nighttime lights time series}.
\newblock
  \bibinfo{howpublished}{\url{https://ngdc.noaa.gov/eog/dmsp/downloadV4composites.html}}
  (\bibinfo{year}{2019}).

\bibitem{gaughan2016spatiotemporal}
\bibinfo{author}{Gaughan, A.~E.} \emph{et~al.}
\newblock \bibinfo{journal}{\bibinfo{title}{Spatiotemporal patterns of
  population in mainland {C}hina, 1990 to 2010}}.
\newblock {\emph{\JournalTitle{Scientific Data}}} \textbf{\bibinfo{volume}{3}},
  \bibinfo{pages}{160005} (\bibinfo{year}{2016}).

\bibitem{donaldson2016view}
\bibinfo{author}{Donaldson, D.} \& \bibinfo{author}{Storeygard, A.}
\newblock \bibinfo{journal}{\bibinfo{title}{The view from above: Applications
  of satellite data in economics}}.
\newblock {\emph{\JournalTitle{Journal of Economic Perspectives}}}
  \textbf{\bibinfo{volume}{30}}, \bibinfo{pages}{171--198}
  (\bibinfo{year}{2016}).

\bibitem{toole2015tracking}
\bibinfo{author}{Toole, J.~L.} \emph{et~al.}
\newblock \bibinfo{journal}{\bibinfo{title}{Tracking employment shocks using
  mobile phone data}}.
\newblock {\emph{\JournalTitle{Journal of The Royal Society Interface}}}
  \textbf{\bibinfo{volume}{12}}, \bibinfo{pages}{20150185}
  (\bibinfo{year}{2015}).

\bibitem{almaatouq2016mobile}
\bibinfo{author}{Almaatouq, A.}, \bibinfo{author}{Prieto-Castrillo, F.} \&
  \bibinfo{author}{Pentland, A.}
\newblock \bibinfo{title}{Mobile communication signatures of unemployment}.
\newblock In \emph{\bibinfo{booktitle}{International Conference on Social
  Informatics}}, \bibinfo{pages}{407--418} (\bibinfo{organization}{Springer},
  \bibinfo{year}{2016}).

\bibitem{e2020uncovering}
\bibinfo{author}{e~Silva, F.~B.} \emph{et~al.}
\newblock \bibinfo{journal}{\bibinfo{title}{Uncovering temporal changes in
  europe’s population density patterns using a data fusion approach}}.
\newblock {\emph{\JournalTitle{Nature Communications}}}
  \textbf{\bibinfo{volume}{11}}, \bibinfo{pages}{1--11} (\bibinfo{year}{2020}).

\bibitem{grantz2020use}
\bibinfo{author}{Grantz, K.~H.} \emph{et~al.}
\newblock \bibinfo{journal}{\bibinfo{title}{The use of mobile phone data to
  inform analysis of covid-19 pandemic epidemiology}}.
\newblock {\emph{\JournalTitle{Nature Communications}}}
  \textbf{\bibinfo{volume}{11}}, \bibinfo{pages}{1--8} (\bibinfo{year}{2020}).

\bibitem{jean2016combining}
\bibinfo{author}{Jean, N.} \emph{et~al.}
\newblock \bibinfo{journal}{\bibinfo{title}{Combining satellite imagery and
  machine learning to predict poverty}}.
\newblock {\emph{\JournalTitle{Science}}} \textbf{\bibinfo{volume}{353}},
  \bibinfo{pages}{790--794} (\bibinfo{year}{2016}).

\bibitem{tianyancha}
\bibinfo{author}{Tianyancha}.
\newblock \bibinfo{title}{{Business Directory}}.
\newblock \bibinfo{howpublished}{\url{https://top.tianyancha.com/companies/}}
  (\bibinfo{year}{2019}).

\bibitem{AIC}
\bibinfo{author}{{The State Administration for Industry and Commerce}}.
\newblock \bibinfo{title}{Analysis on the development of national enterprises
  since the 18th CPC National Congress}.
\newblock
  \bibinfo{howpublished}{\url{http://www.gov.cn/zhuanti/2017-10/27/content_5234848.htm}}
  (\bibinfo{year}{2017}).

\bibitem{de2019geocoding}
\bibinfo{author}{De~Rassenfosse, G.}, \bibinfo{author}{Kozak, J.} \&
  \bibinfo{author}{Seliger, F.}
\newblock \bibinfo{journal}{\bibinfo{title}{Geocoding of worldwide patent
  data}}.
\newblock {\emph{\JournalTitle{Scientific Data}}} \textbf{\bibinfo{volume}{6}},
  \bibinfo{pages}{1--15} (\bibinfo{year}{2019}).

\bibitem{citydefination}
\bibinfo{author}{{The Central People's Government of the People's Republic of China}}.
\newblock \bibinfo{title}{Administrative divisions of the People's Republic of
  China}.
\newblock
  \bibinfo{howpublished}{\url{http://www.gov.cn/test/2005-06/15/content_18253.htm}}
  (\bibinfo{year}{2015}).

\bibitem{cpca}
\bibinfo{author}{DQinYuan}.
\newblock \bibinfo{title}{Chinese province city area mapper}.
\newblock
  \bibinfo{howpublished}{\url{https://github.com/DQinYuan/chinese_province_city_area_mapper}}
  (\bibinfo{year}{2019}).

\bibitem{gaode}
\bibinfo{author}{{Amap Open Platform}}.
\newblock \bibinfo{title}{Geocoding}.
\newblock
  \bibinfo{howpublished}{\url{https://lbs.amap.com/api/webservice/guide/api/georegeo}}
  (\bibinfo{year}{2019}).

\bibitem{coord}
\bibinfo{author}{{wandergis}}.
\newblock \bibinfo{title}{Coordinate transform}.
\newblock
  \bibinfo{howpublished}{\url{https://github.com/wandergis/coordtransform}}
  (\bibinfo{year}{2019}).

\bibitem{limpert2001log}
\bibinfo{author}{Limpert, E.}, \bibinfo{author}{Stahel, W.~A.} \&
  \bibinfo{author}{Abbt, M.}
\newblock \bibinfo{journal}{\bibinfo{title}{Log-normal distributions across the
  sciences: Keys and clues}}.
\newblock {\emph{\JournalTitle{BioScience}}} \textbf{\bibinfo{volume}{51}},
  \bibinfo{pages}{341--352} (\bibinfo{year}{2001}).

\bibitem{gillespie2015fitting}
\bibinfo{author}{Gillespie, C.}
\newblock \bibinfo{journal}{\bibinfo{title}{Fitting heavy tailed distributions:
  The powerlaw package}}.
\newblock {\emph{\JournalTitle{Journal of Statistical Software}}}
  \textbf{\bibinfo{volume}{64}} (\bibinfo{year}{2015}).

\bibitem{figshare}
\bibinfo{author}{Dong, L.}, \bibinfo{author}{Yuan, X.}, \bibinfo{author}{Li,
  M.}, \bibinfo{author}{Ratti, C.} \& \bibinfo{author}{Liu, Y.}
\newblock \bibinfo{title}{A gridded establishment dataset as a proxy for
  economic activity in china}.
\newblock \bibinfo{howpublished}{Figshare,
  \url{https://figshare.com/s/0fed1b024b24c666d595}} (\bibinfo{year}{2020}).

\bibitem{baidu}
\bibinfo{author}{{Baidu Maps}}.
\newblock \bibinfo{title}{Geocoding}.
\newblock
  \bibinfo{howpublished}{\url{https://lbsyun.baidu.com/index.php?title=webapi/guide/webservice-geocoding}}
  (\bibinfo{year}{2019}).

\bibitem{bettencourt2007growth}
\bibinfo{author}{Bettencourt, L.~M.}, \bibinfo{author}{Lobo, J.},
  \bibinfo{author}{Helbing, D.}, \bibinfo{author}{K{\"u}hnert, C.} \&
  \bibinfo{author}{West, G.~B.}
\newblock \bibinfo{journal}{\bibinfo{title}{Growth, innovation, scaling, and
  the pace of life in cities}}.
\newblock {\emph{\JournalTitle{Proceedings of the National Academy of
  Sciences}}} \textbf{\bibinfo{volume}{104}}, \bibinfo{pages}{7301--7306}
  (\bibinfo{year}{2007}).

\bibitem{li2017simple}
\bibinfo{author}{Li, R.} \emph{et~al.}
\newblock \bibinfo{journal}{\bibinfo{title}{Simple spatial scaling rules behind
  complex cities}}.
\newblock {\emph{\JournalTitle{Nature Communications}}}
  \textbf{\bibinfo{volume}{8}}, \bibinfo{pages}{1841} (\bibinfo{year}{2017}).

\bibitem{dong2020understanding}
\bibinfo{author}{Dong, L.}, \bibinfo{author}{Huang, Z.},
  \bibinfo{author}{Zhang, J.} \& \bibinfo{author}{Liu, Y.}
\newblock \bibinfo{journal}{\bibinfo{title}{Understanding the mesoscopic
  scaling patterns within cities}}.
\newblock {\emph{\JournalTitle{arXiv preprint arXiv:2001.00311}}}
  (\bibinfo{year}{2020}).

\bibitem{bettencourt2013origins}
\bibinfo{author}{Bettencourt, L.~M.}
\newblock \bibinfo{journal}{\bibinfo{title}{The origins of scaling in cities}}.
\newblock {\emph{\JournalTitle{Science}}} \textbf{\bibinfo{volume}{340}},
  \bibinfo{pages}{1438--1441} (\bibinfo{year}{2013}).

\bibitem{zheng2017birth}
\bibinfo{author}{Zheng, S.}, \bibinfo{author}{Sun, W.}, \bibinfo{author}{Wu,
  J.} \& \bibinfo{author}{Kahn, M.~E.}
\newblock \bibinfo{journal}{\bibinfo{title}{The birth of edge cities in
  {C}hina: Measuring the effects of industrial parks policy}}.
\newblock {\emph{\JournalTitle{Journal of Urban Economics}}}
  \textbf{\bibinfo{volume}{100}}, \bibinfo{pages}{80--103}
  (\bibinfo{year}{2017}).

\bibitem{he2020watering}
\bibinfo{author}{He, G.}, \bibinfo{author}{Wang, S.} \& \bibinfo{author}{Zhang,
  B.}
\newblock \bibinfo{journal}{\bibinfo{title}{Watering down environmental
  regulation in {C}hina}}.
\newblock {\emph{\JournalTitle{Quarterly Journal of Economics}}}
  \textbf{\bibinfo{volume}{135}}, \bibinfo{pages}{2135--2185}
  (\bibinfo{year}{2020}).

\bibitem{spatialjoin}
\bibinfo{author}{ArcGIS}.
\newblock \bibinfo{title}{Spatial join}.
\newblock
  \bibinfo{howpublished}{\url{https://desktop.arcgis.com/en/arcmap/10.3/tools/analysis-toolbox/spatial-join.htm
  }} (\bibinfo{year}{2020}).

\bibitem{lyons2018comparison}
\bibinfo{author}{Lyons, M.~B.}, \bibinfo{author}{Keith, D.~A.},
  \bibinfo{author}{Phinn, S.~R.}, \bibinfo{author}{Mason, T.~J.} \&
  \bibinfo{author}{Elith, J.}
\newblock \bibinfo{journal}{\bibinfo{title}{A comparison of resampling methods
  for remote sensing classification and accuracy assessment}}.
\newblock {\emph{\JournalTitle{Remote Sensing of Environment}}}
  \textbf{\bibinfo{volume}{208}}, \bibinfo{pages}{145--153}
  (\bibinfo{year}{2018}).

\end{thebibliography}

\end{document}